\documentstyle[aps,prd]{revtex}
\renewcommand{\baselinestretch}{1.5}
\begin{document}
\large
\title{Spin foams, causal links and geometry-induced interactions} 
\author{W. Smilga}
\address{Isardamm 135 d, D-82538 Geretsried, Germany}
\address{e-mail: wsmilga@compuserve.com}
\maketitle 
\renewcommand{\baselinestretch}{1.2}

\begin{abstract}
Current theories of particle physics, including the standard model,
are dominated by the paradigm that nature is basically translation
invariant. Deviations from translation invariance are described by
the action of forces. General relativity is based on a different 
paradigm: There is no translation invariance in general. Interaction
is a consequence of the geometry of space-time, formed by the presence
of matter, rather than of forces.

In recent years the formation of space-time on a quantum mechanical 
level, has been intensively studied within the framework of spin foams,
following an old idea from R. Penrose. In this connection it would be 
appropriate to reconsider the meaning of those paradigms and attempt 
to apply the paradigm of general relativity to particle physics.

A spin foam model with underlying SO(3,2) symmetry is well-suited for 
this purpose. It represents a purely geometric model in the sense of 
the second paradigm. By applying perturbative methods, starting from a
translation invariant first approximation, this model is reformulated
in the sense of the first paradigm. It will be shown that the model then 
defines a space-time manifold equipped with a particle theory in the form 
of locally interacting quantized fields. This includes all four types of
interaction: electromagnetic, weak, chromodynamics and gravitation
together with realistic numerical values of the corresponding coupling 
constants. 
\end{abstract}

\pacs{04.60.Pp, 12.60.-i, 12.20.-m}

\renewcommand{\baselinestretch}{1.0}

\section{Introduction}

Since Isaac Newton formulated his laws of motion, translation 
invariance has become and still is the dominant paradigm of our 
understanding of nature.
Translation invariance is generally understood as a basic symmetry, 
that is more or less disturbed by the action of forces within a
manybody system.
Of course, such a system as a whole is usually assumed not to be under the
influence of external forces. Thus translation invariance is kept up for 
the system as a whole.
Today, this paradigm, together with Lorentz invariance, is the basis 
for the field theoretical formulation of the standard model of elementary 
particles. 

It has been criticized that Newton's laws are a definition of the notion of 
force rather than a law of nature. 
This objection can be formulated as an alternative paradigm: \begin{itshape}
Nature is basically not translation invariant, but translation invariance 
is a useful starting point in developing a physical theory. 
In refining the theory corrective terms have to be added that are defined 
by the differences between real and fictitious geometry. 
These corrective terms are expressed by the concept of `forces'. 
\end{itshape}

The well-known derivation of Newton's theory of gravitation from Einstein's 
theory of general relativity, demonstrates in an impressive way how a theory
based on the second paradigm can be converted into a formulation that is 
in agreement with the first paradigm.
The derivation extracts, from the pure geometric concept of Riemannian 
space-time, gravitational forces as (an approximate) description of the 
same kinematical situation, viewed from a flat coordinate system, rather than 
from curved space-time.

With this beautiful example in mind, it seems natural to ask whether such a 
duality of concepts can be established also in particle physics. 
This article, therefore, takes up the following question: 
Can we find a geometry, not of space-time but of a Hilbert space with the
following properties? It should allow to derive Minkowskian space-time as an 
approximate space-time manifold. At the same time it should deliver interaction
terms, when we use the obtained approximate Minkowskian space-time to formulate 
kinematical relations.

In a way, string theories have been trying to answer similar questions for 
the last three decades. 
But they still have severe difficulties to relate their mathematical models to 
empirical particle physics, and there is no indication that their problems can 
be solved in the near future.

The concepts of spin networks and spin foams, respectively, are also aimed at 
the same question. 
Spin networks were introduced by R.~Penrose \cite{rp} more than 30 years ago, 
in an attempt to describe the geometry of space-time 
in a purely combinatorial way. In recent years spin networks and spin foams 
have gained increased interest as an instrument in the formulation of 
hypothetical structures of space-time at Planck scales aiming at a 
consistent quantum geometry of space-time. 
(See for example J. C. Baez \cite{jcb} for an introduction to spin networks 
and spin foam models and references given therein.) 

At Planck scales, corresponding to a characteristic length of 
$l_P = 1.6 \times 10^{-35}$~m, the usual continuous space-time manifold is 
assumed to break down and have to be replaced by a discrete structure. 
A continuous space-time manifold, equipped with reasonable physics at 
experimentally accessible scales, is then expected as the result of 
a proper large-scale approximation to a spin-network.  
However, in spite of the progress made in recent years it is still unknown 
how such an approximation can be formulated.

A concept of particle physics that is based on `events' in space and time,
connected by `causal links', has been advocated by R. Haag \cite{rh1},
\cite{rh2}. Along this line the comparatively new concept of 
\begin{em}causal\end{em} spin networks has been developed by 
F. Markopoulou and L. Smolin \cite{fmls},\cite{fm}.
It has been used to study the formation of space-time using causal 
links to describe  `histories' within spin networks at Planck scales. 
This concept will be helpful also at `normal' scales, for the interpretation 
of the results that we will obtain.

Despite the unsolved questions of spin networks at Planck scales, we will 
base our considerations on a spin foam concept but we will avoid the 
problems caused by the attempt to find an access to spin foams at Planck 
scales.
Our approach will lead us, in a natural way, to space-time without the need 
to start from Planck scales.

Our strategy will be to make use of comparatively modest basic assumptions
that are closer to well-established concepts of particle physics. 
At first sight these assumptions may appear as too narrow to permit 
a reasonable answer to the above question. 
It will turn out, however, that they yield just enough properties to 
formulate a `tight fitting' realistic particle theory. 
On the other hand their simplicity and the close relationship to conventional 
and well-understood procedures will allow, a mathematically transparent 
transition, to empirical particle physics as formulated by the standard model.

\section{Overview}

We will proceed according to the following programme:

Step 1. Find a suitable symmetry group, to be used within a spin network, that 
contains the homogeneous Lorentz group but not translations as a subgroup.

Step 2. Approximate, in a proper way, this symmetry group by the Poincar\'{e} 
group. 

Step 3. Identify the differences between exact symmetry group and Poincar\'{e} 
group.

Step 4. Bring these differences into a form that can be compared with familiar 
descriptions, preferably with `interaction terms' of the standard model.

Step 5. Compare these interaction terms with those of the standard model, 
if existent, and identify the types of forces that are defined thereby.

A good candidate for such a basic symmetry group is the well-known de~Sitter 
group SO(3,2), which has been used mainly in cosmological models throughout 
many decades. 
It does not contain translations as a subgroup but does contain the Lorentz 
group as a subgroup and can be approximated by the Poincar\'{e} group with 
help of the method of group contraction.

The following study is based on a multiparticle system, defined as a 
spin foam model constructed from spin-1/2 representations of the de~Sitter 
group SO(3,2). Let $\mathcal{H}$ be the Hilbert space of this
multiparticle system.
Group contraction then yields a `tangential' Minkowskian space-time manifold 
equipped with a multiparticle system of non-interacting massive lepton-like 
particles, or `Dirac' particles for short. Let $\mathcal{H}_O$ be its Hilbert 
space.
This will realize steps~1 through 2.

According to step~3, several corrective terms of different structure will be 
identified. 
These terms will be used later to refine the approximation obtained by group 
contraction.
These refinements will lead us, step by step, from the 
contraction limit back towards the exact symmetry of SO(3,2). 
But in doing so, and this is crucial for our approach, we will retain the 
translation invariant $\mathcal{H}_O$ as the mathematical basis of the 
description. 

In order to compare the corrective terms with the interaction terms of the
standard model, we will treat the corrections as a perturbation to the 
(Poincar\'{e} invariant) multiparticle system defined in $\mathcal{H}_O$, 
using conventional Fock space methods. 

Making these corrective terms suitable for a perturbation treatment
will require their linearization by the introduction of auxiliary quantized
fields that act as relativistic potentials.

This will bring one of these corrections into a form, identical to 
the interaction term of the familiar perturbative formulation of quantum 
electrodynamics (QED).
 
In contrast to the standard model, the SO(3,2) model does not contain free 
parameters that can be adjusted to the experimental values of coupling 
constants. 
Therefore, the coupling constants are uniquely determined by the model. 
Their evaluation should either support or fault the suspected relation to
known interactions. 
For this purpose an estimate for the coupling constant for the QED-like 
interaction term based on the SO(3,2) model is derived. 
The estimate reproduces Wyler's heuristic formula \cite{aw}, which is known
to deliver the value of the fine structure constant with a high degree of 
precision. 

The second corrective term cannot be evaluated in the same way, simply because 
the standard model does not contain this type of interaction. 
This interaction is characterized by causing a curving of the space-time 
manifold, obtained before by group contraction. 
This curving is proportional to the distribution of matter, but
does not depend on internal quantum numbers. 
Based on covariance arguments, it can be concluded that a classical 
correspondence limit to this interaction, must result in field equations of 
the type of general relativity. 
So we have good reasons to regard this interaction as a form of quantum 
gravity.

These two contributions do not cover all corrections that have to be applied
to follow the way back to the exact SO(3,2) symmetry. 
We will find which correction defines weak interaction and what
the model can tell us about the nature of neutrinos.
Finally, we will identify quark-like states that give rise to an interaction 
of the type of quantum chromodynamics.

To avoid misunderstandings, we will {\itshape{not}} use the de~Sitter group 
to describe a symmetry of space-time similar to the cosmological model of 
de~Sitter space.
Instead we will use the de~Sitter group as a symmetry group of a Hilbert 
space without any regard to possible consequences for a space-time structure 
in the large or small. Actually, we will not even assume the existence of a 
predetermined space-time continuum.

\section{SO(3,2) based Spin Foam Model }

Given a Lie group $G$. Then a spin network is defined by the following 
properties (see e.g. \cite{jcb}): \newline

\textbf{Definition:} 
\begin{itshape}
A spin network is a triple $\Psi = (\gamma, \rho, \tau)$
consisting of: \newline 

1. a graph $\gamma$, \newline

2. for each edge of $\gamma$, an irreducible representation $\rho_e$ of G,
\newline

3. for each vertex $v$ of $\gamma$, an intertwining operator
\begin{equation}
  \tau_v: \rho_{e_1} \otimes \cdots \otimes \rho_{e_n} \rightarrow 
\rho_{e'_1} \otimes \cdots \otimes \rho_{e'_{n'}} \nonumber
\end{equation}
where $e_1, \ldots , e_n$ are edges incoming to $v$ and 
$e'_1, \ldots , e'_{n'}$ are the edges outgoing from $v$.
\end{itshape} \newline  

Notice that this is the definition of an abstract spin network, not embedded
in any space-time manifold.

Spin networks are general frameworks that have to be filled with contents.
We intend to formulate a model, where $G = SO(3,2)$ and identical irreducible 
spin-1/2 representations of the SO(3,2) are attached to each edge. 

Let $l_{ab}$, $a,b = 0,1,2,3,4$, be representations of the infinitesimal 
generators of SO(3,2) by operators in a quantum mechanical state space  
and let $l_{\mu4}$ be those operators that in the contraction limit converge 
towards the momentum operators $p_\mu$ of the Poincar\'{e} group. 
Then $l_{\mu\nu}$ are the operators of the Lorentz subgroup with 
$\mu,\nu = 0,1,2,3$ with the following commutation relations 
\begin{equation}
[l_{\mu\nu}, l_{\rho\sigma}] = 
-i[g_{\mu\rho} l_{\nu\sigma} - g_{\mu\sigma} 
l_{\nu\rho} + g_{\nu\sigma} l_{\mu\rho} 
- g_{\nu\rho} l_{\mu\sigma}] \mbox{ . } \label{1-a}
\end{equation}
with $g_{\mu\nu}=$ diag $(+1,-1,-1,-1)$.
The operators $l_{\mu4}$ satisfy these commutation relations
\begin{equation}
[l_{\mu4}, l_{\nu4}] = -i g_{44} l_{\mu\nu}
\mbox{     with   } g_{44} = +1       \label{1-b}
\end{equation}
and
\begin{equation}
[l_{\mu\nu}, l_{\rho4}] 
= i[g_{\nu\rho} l_{\mu4} - g_{\mu\rho} l_{\nu4}]  . \label{1-c}	 
\end{equation}

To represent spin we take advantage of the fact that the 4x4-matrices
$s_{\mu\nu}$ and $s_{\mu4}$, built from Dirac matrices, 
\begin{equation}
s_{\mu\nu} :=\, \case{1}{2} \sigma_{\mu\nu}             
\mbox{   and  }  
s_{\mu4} :=\, \case{1}{2} \gamma_\mu                     \label{1-d}
\end{equation} 
with
\begin{equation}
\sigma_{\mu\nu} = \case{i}{2} [\gamma_\mu, \gamma_\nu]   \label{1-e}	
\end{equation} 
and
\begin{equation}
\{\gamma_\mu, \gamma_\nu \} = 2 g_{\mu\nu}               \label{1-f}
\end{equation} 
satisfy the commutation relations of the SO(3,2).  \\
We also will need
\begin{equation}
\gamma_5 = i \gamma_0 \gamma_1 \gamma_2 \gamma_3,        \label{1-g} 
\end{equation}
which anticommutes with all $\gamma_\mu$.
In the Weyl representation $\gamma_5$ is diagonal
\begin{equation}
\gamma_5 = \left(  \begin{array}{rr} 
I & 0 \\
0 & -I                                                   
\end{array} \right) .                                     \label{1-g}
\end{equation}

A Spin-1/2 representation is obtained from the direct product of an 
infinite-dimensional `orbital' part and a finite-dimensional `spin'-part.  
The operators of this representation are given by
\begin{equation}
j_{ab} := l_{ab} + s_{ab}   .                               \label{1-h}
\end{equation}

The motivation for using such a representation is that in the (massive) 
contraction limit (see next section) it reduces to a well-understood 
representation of the Poincar\'{e} group. This describes Dirac particles with 
a given mass $m$ in Minkowski space-time. This will give the model a chance
to be realistic in this respect.
For the moment we can assume that at all vertices $n = n' = 2$.
At each vertex two particles are allowed to exchange their orbital and, 
possibly, spin quantum numbers with an amplitude that will have to be 
determined. 
By the end of this article we will have collected enough information to come
to a precise definition of the vertices.

When we perform the contraction limit, the model will exhibit three space-like 
and one time-like dimension. Therefore, on a time cut the SO(3,2) spin network 
defines a `quantum 3-geometry'. The inclusion of the time-like dimension 
extends the spin network to a spin foam with a `quantum 4-geometry'. 
The vertices of the spin foam will be regarded as `events' in space and time,
which refer to a sharp position in a space-time manifold still to be
defined.

The author considers the formulation of this SO(3,2) model as an intermediate
step on the way to a model that is finally based on 
\begin{em}finite\end{em}-dimensional spin representations of SO(3,2). 
In this article, however, no attempt has been made to extend the model in
this direction.

Instead of using the notions of spin network and spin foams, we could simply
talk about multiparticle systems, since we intend to use the former notions 
in the sense of the latter. The author has decided to use the `modern' 
notions, firstly because these are commonly connected to what this article 
aims at, namely to derive space-time and interactions from a purely geometric 
concept.
Secondly, because the article will clarify some aspects of spin foams.

\section{Group contraction}

The method of group contraction was mathematically formulated by 
E.~In\"on\"u and E.~P.~Wigner \cite{iw} in 1953. By group contraction  
the Poincar\'{e} group is obtained as ``in some sense, a limiting case" 
of the de~Sitter group if the ``de~Sitter radius" $R$ approaches 
infinity.
  
When these authors formulated the contraction limit, they had in mind 
the cosmological model of de~Sitter space-time. 
Our interest is merely to obtain a suitable Poincar\'{e} invariant 
mathematical basis for a following perturbative expansion. 

Group contraction is defined as a restriction of the operators to a
domain of the Hilbert space, where the expectation values of $l_{\mu4}$ are 
large compared to those of $l_{\mu\nu}$, so that for absolute values of 
amplitudes between states $\phi$ and $\phi'$ of this domain the following 
relation holds
\begin{equation}
|\langle\phi|l_{\mu4}|\phi'\rangle| \;\; \gg \;\; 
|\langle\phi|l_{\nu\rho}|\phi'\rangle| 
\mbox{ for all }\mu, \nu, \rho. \label{1-1}
\end{equation}
As a consequence of the commutation relation (\ref{1-b}), the operators 
$l_{\mu4}$ can then be approximated by commuting operators $p_\mu$ that are 
identified as the translation operators of the Poincar\'{e} group $P(3,1)$. 

E.~In\"on\"u and E.~P.~Wigner \cite{iw}, see also F. G\"ursey \cite {fg}, 
have formalized relation (\ref{1-1}) by rescaling the operators $l_{\mu4}$ 
by a factor $1/R$
\begin{equation}
\Pi_\mu = \frac{1}{R}\;l_{\mu4}  \label{1-2}
\end{equation}
and defining Poincar\'{e} momentum by the limit
\begin{equation}
p_\mu = \lim_{R\to\infty} \Pi_\mu.  \label{1-3}
\end{equation}
With these definitions the commutation relations of $l_{\mu4}$ are
\begin{equation}
[ \Pi_\mu, \Pi_\nu ] = \frac{-i}{R^2} l_{\mu\nu}  \label{1-4}
\end{equation}and, therefore,
\begin{equation}
[ p_\mu, p_\nu ] = 0.   \label{1-5}
\end{equation}

F. G\"ursey \cite{fg} has formulated the analogue of eq.\ (\ref{1-2}) 
for spin-1/2 representations
\begin{equation}
\Pi_\mu = \frac{1}{R} l_{\mu4} + \frac{1}{2R} \gamma_\mu .  \label{1-6}
\end{equation} 
(See also C. Fr{\o}nsdal et al\cite{ff}.)
The $\gamma$-term was named ``momentum spin" by F. G\"ursey.

For $R\to\infty$ it is assumed that the operators $l_{\mu4}$ grows 
proportionally to $R$ so that the $l_{\mu4}$ term in (\ref{1-6}) remains 
finite and converges to $p_\mu$. 
The $\gamma_\mu$ term then becomes a second order and can be neglected if 
$R\to\infty$.

By performing the contraction limit we have obtained the algebra
of the Poincar\'{e} group as a kind of high energy approximation to the 
algebra of SO(3,2). In this approximation the multiparticle system, on which 
we base our analysis, is replaced by a system of Dirac particles with 
Hilbert space $\mathcal{H}_O$.

The scaling factor $R$ serves a twofold purpose. It ensures that relation
(\ref{1-1}) is satisfied and it gives the momentum operator the dimension of
an inverse length, if $R$ is understood as the `radius' of de~Sitter 
space-time.

Since our model is not related to de~Sitter space-time we will not make 
further use of $R$, but replace this factor by $1$. 
This means that we will treat the operator $p_\mu$ like $l_{\mu\nu}$ as 
dimensionless. Its eigenvalues are pure numbers. 
In the following, instead of using a scaling factor, we will make use of 
relation (\ref{1-1}), which serves the same purpose.
More about dimensions in the following section.
 
After dropping the factor $1/R$ we can rewrite (\ref{1-6}) in the form 
\begin{equation}
\Pi_\mu =  l_{\mu4} +  \case{1}{2}\gamma_\mu .  \label{1-7}
\end{equation}

It is well-known that from momentum eigenstates $|\mathbf{p}\rangle$,  
${\mathbf{p}} = (p_1, p_2, p_3)$, localized states $|{\mathbf{x}},t\rangle$ 
can be constructed that are 
eigenstates of a `position' operator $X_k, k=1,\ldots,3$,
\begin{equation}
|{\mathbf{x}},t\rangle := 
(2\pi)^{-3/2} \int d^3p \, e^{ipx} |\mathbf{p}\rangle  
\end{equation}
with
\begin{equation}
X_k \, |{\mathbf{x}},t\rangle = x_k \, |{\mathbf{x}},t\rangle .
\end{equation}

The parameter space of ${\mathbf{x}},t$ has then the structure of 
4-dimensional Minkowskian space-time. 
Thus, formally, space-time in our model is generated in a very unspectacular 
way, as an approximate extract of the geometrical structure of the exact 
Hilbert space $\mathcal{H}$. 
Below, we will give this formal definition a precise physical meaning.

\section{Dimension or not dimension}

Physical quantities like length or momentum usually have a `dimension', which
means that their measurement is based on an instruction how to compare the
quantity to be measured with a given standard. The result of the measurement 
is still a pure number, but information is added to this number by which means
it was obtained.

It may, therefore, cause some confusion that we have introduced momentum as a 
dimensionless quantity. 

What does this mean? Actually, we intend to do nothing other than described 
before. To determine the momentum of a particle in tangential space-time, we 
can count the nodes of its wave function within a given area of space-time.
The number of nodes is then proportional to the momentum quantum numbers.
Counting nodes requires an agreement over which area or length, 
respectively, we have to perform the counting. This gives the momentum a
`dimension' in the same sense as stated before.

The other way round, we can also measure a length in tangential space-time, by 
comparing it with an agreed number of nodes of a given momentum standard. 
The reader will realize that this instruction is used in the modern 
definition of the meter by a certain number of krypton wavelengths.

So much for how dimensions can be attached to `dimensionless' quantities.

\section{Corrective terms}

We now start to reconstruct the exact algebra of SO(3,2) from the algebra
of the Poincar\'{e} group. Throughout this paper we assume that relation 
(\ref{1-1}) is valid. 
In the first step we add the term $\case{1}{2}\gamma_\mu$ to the 
momentum operator. 
We then obtain a corrected operator in the form
\begin{equation}
t_\mu :=  p_\mu +  \case{1}{2}\gamma_\mu.   \label{1-8}
\end{equation}

If we add another correction given by
\begin{equation}
c_\mu :=  l_{\mu4} - p_\mu ,   \label{1-9}
\end{equation}
we will have fully reconstructed the original operator $j_{\mu4}$ of SO(3,2).
Since both corrective terms have a significantly different structure it makes 
sense to analyze their contributions separately. 
So let us first try to understand the implications of the spin part and come 
back to the term $c_\mu$ later.

Since in (\ref{1-8}) the spin part is small compared to the orbital part and 
a well-known theory is available for the orbital part, a perturbational
approach is indicated, which treats the spin part as a small perturbation to
a conventional Dirac theory.
In the following we will develop such a perturbational access to the physical 
implications of the operator (\ref{1-8}).

\section{Constraints in multiparticle systems}

Consider the operator $j_{ab} j^{ab}$, $a,b = 0,1,2,3,4$ (summation over
$a,b$), where $j_{ab}$ are given by(\ref{1-h}). 
This is one of the invariant operators of the SO(3,2) algebra, which means 
that it commutes with all $j_{ab}$.

In a multiparticle system we deal with operators
\begin{equation}
J_{ab} := j_{ab} + j'_{ab} + j''_{ab} + j'''_{ab} + \cdots \label{1-10}
\end{equation}
that are defined as the sum of particle-individual operators.
If this system is isolated, which means that it can be described by an 
irreducible representation of SO(3,2), then the invariant operator 
$J_{ab} J^{ab}$ can be assigned a fixed c-number. 

For our spin foam, or any isolated subsystem of it, the evaluation of 
the relation
\begin{equation}
J_{ab} J^{ab} = const                          \label{1-11}
\end{equation}
can, therefore, lead to `selection rules' for possible `transitions' 
at its vertices.

In the following we will approximately evaluate the multiparticle 
operator relation (\ref{1-11}) by using (\ref{1-8}) as an approximation 
to $j_{\mu4}$ and taking (\ref{1-1}) into account, collecting terms of 
the magnitude $p^2$ and $p$ and ignoring terms of lower magnitude. 
We then obtain the expression (written out for two particles with 
momentum $p$ and $p'$)  
\begin{eqnarray}
& & p_\mu p^\mu + 2 p_\mu p'^\mu +  p'_\mu p'^\mu \nonumber \\
& & + \case{1}{2}(\gamma_\mu p^\mu + \gamma_\mu p'^\mu + \gamma'_\mu 
p^\mu + \gamma'_\mu p'^\mu) \nonumber \\
& & + \cdots = const , 				\label{2-1}
\end{eqnarray}
which can be written in the form 
\begin{equation}
T_{\mu} T^{\mu} = \mbox{const.} 	\label{2-3}
\end{equation}
with
\begin{equation}
T_\mu := t_\mu + t'_\mu + \cdots  ,  	         \label{2-2}
\end{equation}
where $t_\mu, t'_\mu, \cdots$ are the particle-individual operators 
given by (\ref{1-8}).

In other words, (\ref{2-3}) defines a constant-of-motion. 
Constants-of-motion have always been useful means to study the internal 
kinematics of a physical system. 

In the Hilbert space $\mathcal{H}_O$, which will be used below for a 
perturbational treatment of (\ref{2-3}), there is another constant operator
\begin{equation}
P_{\mu} P^{\mu} = \mbox{const.}, 
\,\,\,\mbox{where} \,\,\, P_\mu := p_\mu + p'_\mu + \cdots .	
\end{equation} 
In $\mathcal{H}_O$ $P_\mu$ commutes with $T_\mu$ . 
So the modulus $P^\mu P_\mu$ is also a constant with respect to any 
transformation that is generated by $P_\mu$, $T_\mu$ or any generator of the 
homogeneous Lorentz subgroup. 

The fact that both $P^2$ and $T^2$ are constants-of-motion does not mean 
that both represent invariant operators in the sense of representation theory. 
This is rather a consequence of the perturbation algorithm: 
The c-number value of $P^2$ defines the Hilbert space $\mathcal{H}_O$ 
that is used to evaluate $T^2$. It is a c-number by construction. 
The c-number value of $T^2$ really defines a constraint.

The constancy of $P^2$ enables us, to separate (\ref{2-1}) into contributions 
that are quadratic in $p_\mu$ and those that are linear. The latter, therefore, 
form another constant expression 
\begin{eqnarray}
& & \gamma_\mu  (p^\mu  + p'^\mu + \cdots)  \nonumber \\ 	\label{2-4}
&+& \gamma'_\mu (p'^\mu + p^\mu  + \cdots)  \nonumber \\
&+& \cdots 				    \nonumber \\ 				
&=& \mbox{const.} . 			
\end{eqnarray}

Here we find terms that represent the operators of a Dirac equation of 
individual particles, and other terms - like $\gamma_\mu p'^\mu$ - 
that provide for a connection between pairs of particles. 
Whereas the former belong to a Poincar\'{e} covariant
description of a multiparticle system in Minkowski space-time, the latter
can be understood as a perturbation to the Poincar\'{e} covariant system.
So we rearrange the terms
\begin{eqnarray}
& &  \gamma_\mu  (p^\mu  + a^\mu) \nonumber \\			\label{2-5}
&+&  \mbox{similar terms for the other particles} \nonumber \\
&=&  \mbox{const.} 
\end{eqnarray}
with the perturbation term
\begin{equation}
a^\mu = \sum p'^\mu \,\,\, 
\mbox{(sum over all particles except the first)}.  \label{2-6}
\end{equation}
$a^\mu$ describes one of the differences between Poincar\'{e} and exact 
symmetry. With this we have realized step~3 of our programme for the first 
corrective term.
Next we will bring this result into a form that can be compared with 
familiar formulations of multiparticle physics.

\section{Structure of two-particle states }

Since the perturbation term $\gamma_\mu a^\mu$ is basically a sum over
two-particle operators, we will have to make use of two-particle states. 
So let us spend a short look at their general structure (ignoring spin 
variables).

Let 
\begin{equation}
|{\mathbf{P}}\rangle = \int \frac{d^3p}{p_0}\, \frac{d^3p'}{p'_0}\, 
C(\mathbf{p,p'})\, |\mathbf{ p, p'} \rangle   \label{3-1}
\end{equation} 
be a two-particle state with 4-momentum $P=({\mathbf{P}},P_0)$ of a 
two-particle representation 
of P(3,1) with $P^2 = M^2$ in a state space ${\mathcal{H}}_M$. 
The two-particle states $|\mathbf{p, p'}\rangle$ belong to the direct 
product of one-particle states $|\mathbf{p}\rangle$ with 4-momentum 
$p=({\mathbf{p}},p_0)$ . 

With $p = k - q$, $p' = k + q$, $2k = P$ we can rewrite
\begin{equation}
|{\mathbf{P}}\rangle = \int \frac{d^3q}{q_0}\, \, \tilde{C}(\mathbf{q}) \,  
 |\mathbf{k - q, k + q} \rangle . 	\label{eq32}
\end{equation}
From
\begin{equation}
p^2 = p'^2 = m^2,  		\label{3-2}
\end{equation} 
where $m$ is the particle mass, and
\begin{equation}
P^2 = (p + p')^2 = M^2   \label{3-3}
\end{equation} 
we obtain
\begin{equation}
q^2 = m^2 - \case{1}{4}M^2,  		\label{3-4}
\end{equation} 
\begin{equation}
q_0^2 = m^2 - \case{1}{4}M^2 + {\mathbf{q}}^2  \label{3-5}
\end{equation} 
and 
\begin{equation}
kq = 0 . 			\label{3-6}
\end{equation} 

Conditions (\ref{3-4}) through (\ref{3-6}) express the fact that 
$|{\mathbf{P}}\rangle$ is a state of an two-particle representation 
characterized by the total mass $M$.

We can formulate the state (\ref{3-1}) also in terms of wave functions
\begin{equation}
e^{-iPx} = \int \frac{d^3q}{q_0}\, \, \tilde{C}({\mathbf{q}}) 
\,\, e^{-i(k-q)x} \, e^{-i(k+q)x} ,  \label{3-7}
\end{equation}
which are obtained by formally multiplying the ket-states 
$|{\mathbf{P}}\rangle$, $|{\mathbf{k-q}}\rangle$ and $|{\mathbf{k+q}}\rangle$
by its associated bra-states $\langle x|$.

The momenta $p$ and $p'$ of each term under the integral in (\ref{3-1}) 
adds up to $P$. In other words, the individual momenta are `entangled' within 
two-particle states.

\section{Interaction term in Fock space }	 

Let us now return to our multiparticle Hilbert space $\mathcal{H}_O$, 
which has been defined by the direct product of one-particle state spaces, 
which in turn are defined by momentum eigenstates that satisfy the Dirac 
equation. We reformulate $\mathcal{H}_O$ with the help of standard Fock 
space methods. 

The `free' part of our system is easily converted into a Fock space 
formulation following the usual `second quantization' of the Dirac field. 
We will skip this step and refer to standard textbooks (see e.g. \cite{gs}).
The field operator of the Dirac field (taken from this reference) has
the form
\begin{equation}
\psi(x) = (2\pi)^{-3/2} \!\!\int\!\! d^3p 
\left( b_s({\mathbf{p}}) u_s ({\mathbf{p}}) e^{-ipx} 
\!+\! {d_s({\mathbf{p}})}^\dagger v_s({\mathbf{p}}) 
e^{ipx} \right).                                        \label{4-1}
\end{equation}
A similar expression defines the Dirac adjoint operator $\bar{\psi}(x)$. 
$b^\dagger_s({\mathbf{p}}), b_s({\mathbf{p}})$ are electron emission 
and absorption operators, $d^\dagger_s({\mathbf{p}}), d_s({\mathbf{p}})$ 
are the corresponding operators for positrons.
They satisfy the usual anticommutation relations of the Dirac field.

As a first attempt we represent our two-particle perturbation terms 
$\gamma_\mu p'^\mu$ in Fock space in the following form  
\begin{equation}
\int\! d^3x\,d^3x' \, \bar{\psi}(x)\gamma_\mu \psi(x) \,\, 
\bar{\psi}(x') p^\mu \psi(x') .                          \label{4-2}
\end{equation}
This Fock space operator is not yet adapted to its immediate insertion into 
a perturbation calculation since it has two major flaws.

Firstly: It is built as a product of two Fock operators. 
This will result in a non-linear equation for $\psi(x)$. 
Such an equation cannot be treated by a standard perturbation algorithm. 
We will find a solution to this problem by a linearization method.

Secondly: Remember that a standard quantum mechanical perturbation 
calculation takes place within the given Hilbert space $\mathcal{H}_O$.
This requires the combination of a perturbation term with a projection 
operator into $\mathcal{H}_O$ (see e.g. \cite{pam}). 
Therefore, we have to incorporate a suitable projection mechanism.
Since the interaction term is a two-particle operator we can base this
mechanism on two-particle subspaces of $\mathcal{H}_O$. 
As shown below, we can implement a suitable projection mechanism 
by collecting only those terms of (\ref{4-2}) that contribute if we evaluate 
this operator for two-particle states with total momentum $P$, $P^2=M^2$, 
$M$=c-number, of a given (irreducible) two-particle state space 
${\mathcal{H}}_M$. 
This will ensure the correct projection onto $\mathcal{H}_O$ also in the 
general case of more than two particles.
One can convince oneself that this procedure is also necessary to ensure
such a projection in the general case.

\section{Implementing a projection onto the Hilbert space}

Consider the following contribution to (\ref{4-2})
\begin{equation}
\dots {\bar{b}({\mathbf{p+k}}) \, \gamma_\mu \, } {b({\mathbf{p}})\,\,} 
{\bar{b} ({\mathbf{p'-k'}}) \, p^\mu} \, b(\mathbf{p'})\dots   \label{4-3}
\end{equation}
(We have omitted the factors of $u_s$ and $v_s$ for a moment.)
If we evaluate this operator for a two-particle state, then only terms
with $\mathbf{k=k'}$ will be involved, as a consequence of momentum 
entanglement within two-particle states.
 
This is true for every state of ${\mathcal{H}}_M$ with a given momentum $P$. 
And since every two-particle state of ${\mathcal{H}}_M$ can be represented by 
a superposition of two-particle momentum eigenstates, it is generally valid.
So we can drop the restriction to a fixed $P$ and collect all contributions 
that belong to the same $\mathbf{p}$ and $\mathbf{k}$. 
Hence, we can write 
\begin{equation}
\dots {\bar{b}({\mathbf{p + k}}) \, \gamma_\mu \, b({\mathbf{p}})} \, 
a^\mu({\mathbf{k}}) \dots                             \label{4-4}
\end{equation}
with
\begin{equation}
e\,a^\mu({\mathbf{k}}) := \int dV(p')
\,\bar{b}({\mathbf{p' - k}}) \, p^\mu \, b({\mathbf{p'}}),      \label{4-5}
\end{equation}
where $dV(p')$ indicates a summation over all terms that contribute to
a given $\mathbf{k}$. 
As a precaution we have included a normalization factor $e$ into the 
definition of $a^\mu$ that will have to be determined after we have decided 
about the normalization of $a^\mu$. 
An analogous consideration is valid for positron and mixed terms.

\section{Splitting the interaction term}

Let us analyze the meaning of $a^\mu(\mathbf{k})$ in more detail. 
If we evaluate (\ref{4-4}) for a two-particle state, the second particle 
will contribute a complex amplitude given by the expectation value of 
$a^\mu(\mathbf{k})$ that acts as a multiplicative weight to the 
expectation value of the first particle term. 
This weight depends on $\mathbf{k}$ and fully describes the contribution of 
a second particle to the total expectation value. A long as our focus is on 
the first particle, then all we need to know about other particles are the 
complex weights that apply to the expectation values of particle one. 
To keep track of the weighting factors that apply to each `transition' 
$p \rightarrow p+k$ in (\ref{4-4}) we need a suitable `bookkeeping' system.

We can establish such a bookkeeping system by introducing an auxiliary Fock 
space with operators that emit and absorb quanta with momentum $\mathbf{k}$. 
If we prepare a state in this Fock space by applying an emission operator
multiplied by a complex amplitude onto the vacuum state, then a later
application of an absorption operator will redeliver this amplitude. 
This is exactly what we need. 

We have some freedom in doing this, as long as the system is able to keep 
track of the amplitudes of the momenta $\mathbf{k}$.
So let us replace the operator $a^\mu(\mathbf{k})$ of (\ref{4-5}) 
by an operator $A^\mu(\mathbf{k})$ of our bookkeeping system. This replacement
means that now $A^\mu(\mathbf{k})$ does not act on a fermions Fock state but
rather on an `intermediate' state in the bookkeeping Fock space. 
By this trick we have split the interaction term into two parts:
The first term acts on the first particle and places a momentum 
${\mathbf{k}}$ into the bookkeeping registry, the second term takes the same 
momentum from the registry and acts onto the second particle, and vice versa.
Notice that each half of the interaction term now has the form of a 
$\gamma_\mu$ vertex, but from the view of each particle the other one seems 
to have a $p^\mu$ vertex in agreement with the form (\ref{4-3}) 
of the original interaction term.

In this way we have linearized the `equation of motion' for $\psi(x)$ at the 
expense of introducing another quantized field.  
 
In their function as Fock space absorption and emission operators $A_\mu$ 
and its ajoint $A_\mu^\dagger$ have to satisfy the following commutation 
relations
\begin{equation}
[A^\mu({\mathbf{k}}), A^\nu ({\mathbf{k'}})^\dagger] 
= \delta^{\mu\nu}\delta({\mathbf{k - k'}}) .                 \label{4-7}
\end{equation}
Then $A^{\mu\dagger}(\mathbf{k})$ are emission operators and 
$A^\mu(\mathbf{k})$ absorption operators for quanta with momentum 
$\mathbf{k}$. 

We define the following operators known from the conventional formulation
of quantum electrodynamics (see e.g. \cite{gs}) 
\begin{eqnarray}
A^j(x) = (2\pi)^{-3/2} \!\!\int\!\! \frac{d^3k}{k^0 \sqrt{2}} 
\left( A^j({\mathbf{k}}) e^{-ikx} 
+ A^j({\mathbf{k}})^\dagger e^{ikx} \right), \nonumber \\
j=1,2,3,                                                     \label{4-8}
\end{eqnarray}
and
\begin{equation}
A^0(x) = (2\pi)^{-3/2} \!\!\int\!\! \frac{d^3k}{k^0 \sqrt{2}} 
i \left( A^0({\mathbf{k}}) e^{-ikx} 
+ A^0({\mathbf{k}})^\dagger e^{ikx} \right). \nonumber      \label{4-9}
\end{equation}
$k^0$ shall be determined by condition (\ref{3-5}) when these operators are 
evaluated within two-particle states. 
(In the `free radiation field' $k^0$ is `on-shell': $k^0 = |\mathbf{k}|$.)  

Coming back to the expression (\ref{4-4}), we add space-time dependencies 
to the emission and absorption operators, as prescribed by (\ref{3-7}). 
By making use of $A^\mu$ now we obtain
\begin{equation}
\dots \bar{b}({\mathbf{p+k}}) e^{i(p+k)x} \gamma_\mu 
b({\mathbf{p}}) e^{-ipx}\,\, 
A^\mu({\mathbf{k}}) e^{-ikx} \dots .                        \label{4-10}
\end{equation}
Notice that the correct space-time dependency of $A^\mu$ is determined by 
(\ref{4-5}).

After inserting the spin functions $u_s(\mathbf{p})$ and $v_s(\mathbf{p})$
these terms and the corresponding positron and mixed terms add up to a 
Fock operator in the form
\begin{equation}
e \int d^3x\, : \bar{\psi}(x)\gamma_\mu \psi(x) : A^\mu(x), \label{4-11}
\end{equation}
where $::$ stand for normal ordering of emission and absorption operators
(all emission operators stand left of all absorption operators).
This is the form of the interaction term of quantum electrodynamics (QED).  

Actually, we have done something very familiar from classical and  
quantum mechanics: 
We have linearized a two-body problem by introducing a potential that 
describes the action of particle 2 on particle 1 and vice versa. 
This potential has not been obtained by a formal `quantization rule' 
applied to a classical potential, but by explicit construction on the quantum 
mechanical level. This provides us with a full insight into its mathematical 
and physical implications.

The interaction term (\ref{4-11}) uniquely defines the structure of QED.
Therefore, the full machinery of QED is available to analyze the 
effect of this term in a perturbation calculation. The result of this
analysis is well-known and signifies that a Dirac particle within the SO(3,2) 
model shows the properties of an electrically charged particle.

\section{Iterated fields and Feynman graphs}

The explicit construction of the interaction term, makes it possible to give
a well founded interpretation of the mathematical and physical contents of 
the iterated field operator that we obtain from perturbation calculations. 

Since the interaction term is identical to that of QED, the application of 
the perturbation algorithm will produce all those terms that 
are known from QED and that are represented there by Feynman graphs.
Some of these graphs or components of graphs, respectively, have been given 
the interpretation of pair creation, vacuum fluctuation and, not to forget, 
of `free' photons.

Let us start with the notion of photons. The photon field in the SO(3,2) model
has been introduced as a means to describe the exchange of momentum between 
two fermions. A photons connects {\itshape{two}} vertices in the sense
of a causal link. 
Therefore, the existence of a `free' photon is only due to the fact 
that it has been emitted at a point $x$ in space-time {\itshape{and}} 
that it will be absorbed at another point $x'$ - possibly far away from $x$. 
As we know from basic nuclear physics the exchange of massive particles leads 
to short distance interactions. (This information can also be obtained from
the structure of the propagators representing internal photon lines.)
Therefore, it is understandable that only mass zero photons can be exchanged 
over a large distances. 
This means: `free' photons are `on-shell' quanta of the vector potential with 
mass zero.

This interpretation clearly shows that the electromagnetic interaction 
basically is a long range interaction, with photon representing 
\begin{em}causal links\end{em} rather than `real' particles. 
Relativistic causality in this connection is guaranteed by the 
causal properties of the commutation functions.

Nevertheless, the introduction of the photon field allows us to describe the
interaction by a local (point like) electron-photon vertex. 
But keep in mind that the basic interaction process always consists of two
such vertices.

One word about external fermion lines corresponding to `asymptotic states': 
These states in our model simply mean one-particle states in $\mathcal{H}_O$. 
There is no need to complicate matters in defining asymptotic states as 
a limit for $t \to \pm \infty$. 

Now we come to `pair creation'. This process is characterized by a Feynman 
graph where, as a result of a scattering process, there are two outgoing 
fermion lines - an electron line and a positron line. 
Since the Fock operator that represents the interaction term, by construction, 
cannot physically create or annihilate fermions, we obviously have to 
accept the fact that, as a consequence of relativistic covariance, there is 
the possibility of scattering not only in space-like but also in time-like 
directions. Scattering in space-like directions means (in the centre-of-mass 
system) a change of sign of a space-like component of a 4-momentum. 
Scattering in time-like direction correspondingly means a change of sign of 
the time-like component of a 4-momentum.
As in the space-like case, after the scattering event particles change 
their direction of movement, in the time-like case an electron
after the scattering event moves backwards in time and is called a positron
from now on.

The picture of a particle of negative energy running backwards in time is  
Feynman's view of a positron \cite{rf}. 
Here it develops in a most natural way. This `running backwards in
time' is the deeper reason why, for states of negative energy the absorption 
operator in (\ref{4-1}) had to be replaced by an emission operator.

Now to `vacuum fluctuation'. Some Feynman graphs contain closed fermion 
loops that have been interpreted as creation and annihilation of virtual 
electron-positron pairs from the `vacuum'.
`Virtual pairs' have long been considered as a clear indication that a theory 
of elementary particles cannot be established without the introduction of 
relativistic quantized fields, and that such fields theories automatically 
mean a theory with an infinite number of particles. 

What does the SO(3,2) model tell us about `vacuum fluctuation'?

Closed loops are a consequence of the perturbative iteration of a 
Fock operator which, by construction, cannot create fermions from the vacuum. 
If this Fock operator is iterated it still cannot create fermions.
Again, if this Fock operator is applied to two-particle states, there
are exactly two particles involved and not more.
A down-to-earth interpretation of such loops is easily obtained if we
remember that internal lines in a Feynman graph are generated by interchanges
of emission and absorption operators, by using their commutation relations. 
These interchanges essentially deliver delta-functions for the momenta, which, 
within the perturbation algorithm, are extended in a covariant way to 
4-momentum space.
Therefore, internal lines simply keep track of 4-momentum within
the perturbation algorithm and have nothing to do with particles, not even 
with `virtual' ones. This is in contrast to external lines which do represent
one-particle and one-photon states.

In the past the highly imaginative picture of a vacuum swirling with virtual 
pairs, has proven more attractive than a rational analysis of the perturbation 
algorithm. Unfortunately, this picture has influenced our way of looking at 
particle physics throughout five decades.

\section{Gauge invariance }

In the standard formulation of QED an interaction with a `gauge field' is 
introduced by postulating gauge invariance of second kind.
In contrast to this rather formal procedure our evaluation of the SO(3,2) 
model lead us to an explicit construction of the interaction term and the 
`bookkeeping field' from known elements of the electrons Fock space. 
This enables us to {\itshape{prove}} gauge invariance of the second kind 
rather than only postulate it.

Consider the two-particle wave function (\ref{3-7}) and multiply the first 
one-particle wave function on the right side by $e^{i \Lambda(x)}$ and the 
second by $e^{-i \Lambda(x)}$. 
Obviously, this leaves the two-particle wave function invariant. 
Now apply the first term in (\ref{2-5}) with all momentum operators expressed 
by $-i \partial^\mu$ to this `gauge transformed' two-particle state. 
Then $-i \partial^\mu$ applied to the wave function of the first particles 
contributes a term $\partial^\mu \Lambda(x)$ 
whereas $a^\mu$ applied to the wave function of the second particles by making
use of (\ref{2-6}) delivers the same term with opposite sign. 
Both terms cancel each other.
This is exactly what is meant by gauge invariance of the second kind.

\section{Estimate of the coupling constant}

Unlike the standard formulation of QED, where the coupling constant enters as 
a free parameter that has to be determined by the experiment, our approach
does not leave room for any free parameter. This means that the coupling 
constant $e$ is determined by the theory and, therefore, should be calculable.

The coupling constant is defined by the normalization factor $e$ in 
(\ref{4-5}).
After replacing $a^\mu({\mathbf{k}})$ by $A^\mu({\mathbf{k}})$ the 
normalization of the Fock operators in 
(\ref{4-5}) is fixed by their commutation relations.
Then $e$ can be determined by correctly `counting' all contributions to the 
integral - in other words: by a careful analysis of the volume element of 
the integral in (\ref{4-5}). 

More than 30 years ago A. Wyler \cite{aw} discovered that the 
fine-structure constant $\alpha$ can be expressed by volumes of certain 
symmetric spaces. Being a mathematician he was not able to put his 
observation into a convincing physical context. 
Therefore, his work was criticized as fruitless numerology \cite{br}.  

Wyler's idea was picked up later by F. D. Smith, Jr. \cite{fds} who extended 
Wyler's heuristic approach into a general scheme based on a 
fundamental Spin(8) symmetry. 
Smith was then able to express coupling constants and relations of particle 
masses by characteristic volumes with a remarkable degree of precision. 

Let us see how far our model will lead us and whether we possibly can find a 
physical explanation for these authors' observations. 

The following will be oriented to a scattering process 
(M{\o}ller scattering), which means two vertices.
Therefore, the factor $e$ and the volume element of (\ref{4-5}) will enter 
twice into the estimate.

In (\ref{4-5}) we already have parameterized the contributions to the 
interaction term by the parameters ${\mathbf{p'}}$ and ${\mathbf{k}}$. 
The way in which the parameters ${\mathbf{k}}$ are used in the perturbation 
calculation, defines the parameter space of ${\mathbf{k}}$ as (a subspace of) 
the Euclidean $R^3$. 
If we keep ${\mathbf{k}}$ fixed, we are left with the integral over 
${\mathbf{p'}}$ and our task will be to determine the multiplicity or the 
integration volume, respectively, of the contributions with respect to 
${\mathbf{p'}}$. 

The basis for the evaluation of the integration volume is the particle 
momentum and the homogeneous Lorentz group acting on the particle momentum. 
The SO(3,1) acts transitively on a particles mass shell
\begin{equation}
p_0^2 -p_1^2 - p_2^2 - p_3^2 = m^2 . \label{5-1}
\end{equation} 
The independent parameters $p_1,p_2,p_3$ span a 3-dimensional parameter space. 
For a two-particle state of a representation with mass $M$ we have instead
\begin{equation}
(p_0+p_0')^2-(p_1+p_1')^2-(p_2+p_2')^2-(p_3+p_3')^2 = M^2. \label{5-2}
\end{equation}
We can convert this into
\begin{eqnarray}
p_0^{2}+p_0^{'2}-p_1^{2}-p_1^{'2}-p_2^{2}
-p_2^{'2}-p_3^{2}-p_3^{'2}          \nonumber \\
+2p_0 p_0'-2p_1 p_1'-2p_2 p_2'-2p_3 p_3' = M^2.	\label{5-3}
\end{eqnarray}
From (\ref{5-1}) and (\ref{5-2}) follows that
\begin{equation}
p_0 p_0'-2p_1 p_1'-2p_2 p_2'-2p_3 p_3' = \kappa^2   \label{5-4}
\end{equation}
must be invariant.
Therefore,
\begin{eqnarray}
p_0^{2}+p_0^{'2}-p_1^{2}-p_1^{'2}-p_2^{2}-p_2^{'2}-p_3^{2}-p_3^{' 2} 
\nonumber \\
 = M^2 - \kappa^2 .	\label{5-5}
\end{eqnarray}
The symmetry group of this quadratic form is SO(6,2). 
Relation (\ref{5-4}) reduces the number of independent parameters from 6 to 5 
and thereby SO(6,2) to SO(5,2). 
SO(5,2) acts transitively on this 5-dimensional parameter space.
Each point in this parameter space corresponds to a state in the two-particle 
state space ${\mathcal{H}}_M$. Therefore, the volume of the parameter space
delivers a measure for the number of states that can contribute to the
interaction term.

Given a point $Q$ in this parameter space, then other points can be reached 
by applying a linear transformation of SO(5,2) to $Q$. There are certain 
transformations that do \begin{em}not\end{em} change the point $Q$. 
These transformations form the subgroup S(O(5) x O(2)).
This is the isotropy subgroup or stabilizer of $Q$. Therefore, to obtain the
multiplicity of states, we have to start from the coset space 
$D_5$ = SO(5,2)/S(O(5) x O(2)) rather than from SO(5,2).  

$D_5$ is a symmetric space. 
By construction $D_5$ is isomorphic to ${\mathcal{H}}_M$.
It is known from the work of Hua and Lu \cite{hl}
that $D_5$ can be represented by matrices;
that is, this symmetric space is isomorphic to the real hyperball
\begin{equation}
{\mathcal{R}}_R(5,2) = \{X\in R^{5\times2} \,|\, I-XX' > 0 \}. \label{5-6}
\end{equation}
(See \cite{fkklr} for a modern introduction to symmetric spaces.)
Hua \cite{lkh} has calculated volumes of ${\mathcal{R}}_R(5,2)$ and other
domains.
In contrast to ${\mathcal{R}}_R(5,2)$ $D_5$ has an infinite volume.

Consider now a two-particle state $|{\mathbf{P}}\rangle$ with 4-momentum $P$. 
There is another volume associated with $D_5$. 
This is the subspace of all points that correspond to 
a situation where for one of the particles $p_0 = m$ and the other particle 
has reached its maximum value of $p'_0 = P_0 - m$. 
For reasons of symmetry this volume is spherical symmetric and isomorphic to 
the border sphere $C_5$ of $D_5$. $C_5$ has 4 dimensions.
Then all states with given state $P_0$ are confined to a volume $\bar{D}_5$ 
inside of $C_5$ and including $C_5$.
The subspace $\bar{D}_5$ of $D_5$ is finite and can be mapped onto 
${\mathcal{R}}_R(5,2)$ by an isometric mapping. 
 
If $Q = (q_1,...,q_5)$ is a point of $\bar{D}_5$ that is mapped into
a point $S = (s_1,...,s_5)$ of ${\mathcal{R}}_R(5,2)$, then we can 
establish a one-to-one relationship such that 
\begin{equation}
q_i = r \, s_i,    \label{5-7}
\end{equation}
where $r$ is a properly chosen scaling factor. This gives us the choice
to use either $q_i$ in $\bar{D}_5$ or $s_i$ in ${\mathcal{R}}_R(5,2)$
as integration parameters. 

To be consistent with Smith's terminology we will calculate all volumes in 
${\mathcal{R}}_R(5,2)$. We will use the notation $V(D_5)$ for the volume 
that corresponds to $\bar{D}_5$ but is calculated in ${\mathcal{R}}_R(5,2)$ 
and will remember that we have to apply the correct number of scaling 
factors $r$.

$C_5$ has another important property: If particle 1 is initially at rest
and a second particle with a given momentum $p'$ is added to form a
two-particle state, then this state corresponds to a point on $C_5$ as 
described before. Other states can be generated from this `initial' state
by the exchange of momentum. Therefore, to determine all states that 
are eventually involved we have first to collate all initial states. 

This means, we have to perform an integration over $C_5$ with a volume 
element $d^4s/V(C_5)$.
This delivers a first volume factor of $1/V(C_5)$.

To collect all possible momentum changes of particle 1 we have to integrate
over $\bar{D}_5$. From condition (\ref{3-6}) it follows that for a given 
$|{\mathbf{P}}\rangle$ only momentum exchanges in the subspace perpendicular 
to $P$ have to be considered. 
Since the direction of the total momentum is undetermined (when we are 
constructing the interaction operator), we have to keep the integration 
over $D_5$.  
We compensate for this by a volume factor of $1/V(S_4)$ where 
$S_4 = SO(5,2) / SO(4,2) = SO(5) / SO(4)$ is the unit sphere in 4 dimensions. 
This reduces the number of independent parameters to 4. 
Let $(s_1,..,s_4)$ be a new set of independent parameters corresponding to
a new set $(q_1,..,q_4)$.  

If we integrate over $\bar{D}_5$ using this new parameter set, each $s_i$ 
will be responsible for a contribution of $V(D_5)^{1/4}$ to the volume of 
$\bar{D}_5$ .
Three of these parameters can now be mapped onto the transferred momentum 
${\mathbf{k}}$.
The fourth parameter $s_4$, obviously, corresponds to a momentum transfer 
within each of the particle momentum, without any momentum transfer between 
the particles. Such transitions contribute to the volume of $C_5$.
We can perform the integration over $s_4$ and obtain a correcting factor 
to the already calculated volume $V(C_5)$ of $V(D_5)^{1/4}$.

There are three more factors that contribute to the multiplicity of momentum
states. One is related to the spin components of the particle states, which
give each momentum state a multiplicity of $2\pi$ because of the periodicity
of spin states. The other factor is related to the (relative) phases of the 
momentum states within multiparticle states. By adding another factor of 
$2\pi$ we take into account this degree-of-freedom.
Finally, remember that there are two terms $\gamma^\mu p'_\mu$ and 
$\gamma'^\mu p_\mu$ that contribute to the interaction. This delivers a factor
of $2$.

After extracting these constant factors from the integral we are left with an 
integration over the $p'$ parameter space where now the integrand should enter 
with a multiplicity of one within the $p'$-parameter space - provided that we 
have correctly captured all factors that determine any multiplicities. 
Collecting these factors we end up with  
\begin{equation}
8 \pi^2 \,V(D_5)^{1/4} \, / \, (V(S_4) \, V(C_5)).    	\label{5-8}
\end{equation}
This is essentially Wyler's formula.

The volumes $V(D_5)$ and $V(C_5)$ have been calculated by 
L. K. Hua \cite{lkh}. 
$V(S_4)$ is the volume of the unit sphere $S_4$ in 4 dimensions. With
\begin{equation}
V(C_5) = \frac{8 \pi^3}{3},        			\label{5-9}
\end{equation}
\begin{equation}
V(D_5) = \frac{\pi^5}{2^4\, 5!},   			\label{5-10}
\end{equation}
\begin{equation}
V(S_4) = \frac{8 \pi^2}{3}         			\label{5-11}
\end{equation}
we obtain 
\begin{equation}
\frac{9}{8 \pi^3} \left(\frac{\pi^5}{2^4 \, 5!}\right)^{1/4} . 	\label{5-12}
\end{equation}
If we identify this value with the coupling constant 
$e^2/(2\pi)^2 = \alpha/\pi$ in the S-matrix element for M{\o}ller scattering 
we obtain a value for 
$\alpha$
\begin{equation}
\alpha = \frac{9}{8 \pi^4} \left(\frac{\pi^5}{2^4 \, 5!}\right)^{1/4} 
= 1/137.03608245.   				\label{5-13}
\end{equation}
Although intended only as an estimate, this result is in agreement with 
experimental values in five-parts in ten-million. 
(A value of 137.035 999 93(52) has been determined from the magnetic moment 
of the electron \cite{kino}.)

We can easily convince ourselves that the scaling factors $r$ either cancel 
or are absorbed in the volume element $d^3q$.

This result means more than just an estimate of the coupling constant. 
It delivers numerical support of our statement that only states of
the same irreducible representation of the Poincar\'{e} group contribute to 
the interaction term. In other words, it can be considered as 
a confirmation of our implementation of the projection operator into the 
interaction term.

In this section we have obtained a theoretical justification for the scheme 
presented in the article by F. D. Smith, Jr. \cite{fds}. 
Concerning coupling constants of other interactions, the reader is 
referred to Smith's article.

\section{Quantum gravity}

Let us now examine the second corrective term (\ref{1-9}) that has been 
identified in the beginning.
Consider the representation of the generators of the 
SO(3,2) in the following form
\begin{equation}
l_{ab} = x_b p_a - x_a p_b,                            \label{7-1}
\end{equation}
where the `momentum' operators $p_a$ are represented by differential 
operators $-i\partial/\partial x^a$.
In writing down this expression we have embedded the tangential space-time 
into a 5-dimensional pseudo-Euclidean space by adding $x_4$ as an additional
(time-like) coordinate. 

Within a neighborhood $\mathcal{N}$ of the point $x_\mu=0, x_4=1$, group 
contraction approximates the operators 
\begin{equation}
l_{\mu4} = x_4 p_\mu - x_\mu p_4 , \,\,\,\,\, \mu=0,\ldots,3,   \label{7-2}
\end{equation}
by momentum operators $p_\mu$.

A better approximation is obtained if the second term on the right of 
(\ref{7-2})   
\begin{equation}
c_\mu = - x_\mu p_4                                              \label{7-3}
\end{equation}
is added.
This term delivers a contribution to the difference between translations 
and exact SO(3,2) operators that is of the first order in $x$ within the
neighborhood $\mathcal{N}$. 
This operator has the structure of a `translation' operator in the direction 
of $x_4$ weighted with $x_\mu$. 
So we can say that this corrective term leads out of the tangential 
space-time into the direction of $x_4$ that is perpendicular to tangential 
space-time. 
Or, in other words, it adds curvature to the originally flat space-time.

Let us see how this generation of curvature depends on the properties of a 
two-particle state. 
As we have proceeded before, we start from a `constant-of-motion' 
given by the constant operator
\begin{equation}
L^{ab}L_{ab} = (l^{ab} + l'^{ab})(l_{ab} + l'_{ab}) .  \label{7-4}
\end{equation}

Again we assume that within the neighborhood $\mathcal{N}$ the relations 
(\ref{1-1}) is valid. 
If we evaluate the product (\ref{7-4}) with $l_{\mu4}$ approximated by
\begin{equation}
t_\mu:= p_\mu + c_\mu                                      \label{7-5}
\end{equation}
and take only terms that are quadratic and linear in $p$, 
then within the linear terms we obtain mixed terms of operators of 
particles 1 and 2 in the form of $p_\mu c'^\mu$ and 
$p'_\mu c^\mu$.
These terms define a correction to those terms that we would have
obtained also in a pure Poincar\'{e} invariant situation.

If we evaluate the mixed terms for a two particle state with well-defined 
total energy-momentum $P$ we obtain contributions in the form
\begin{equation}
\langle\phi({\mathbf{p}} + {\mathbf{k}})|\,p_\mu|\phi({\mathbf{p}})\rangle 
\langle\phi({\mathbf{p}}' - {\mathbf{k}}')|\,c'^\mu|\phi({\mathbf{p}}')\rangle+
\langle\phi({\mathbf{p}} + {\mathbf{k}})|\,c^\mu|\phi({\mathbf{p}})\rangle 
\langle\phi({\mathbf{p}}' - {\mathbf{k}}')|\,p'_\mu|\phi({\mathbf{p}}')\rangle.   
\label{7-6}
\end{equation}
Since in the total momentum must be constant, we have 
${\mathbf{k}} = {\mathbf{k}}'$. 

From this we can directly read that the curvature generating correction,
to the kinematics of particle 2 caused by 
$\langle\phi({\mathbf{p}}' - {\mathbf{k}})|\,c'^\mu|\phi({\mathbf{p}}')\rangle$,
is weighted by the 4-momentum amplitude 
$\langle\phi({\mathbf{p}} + {\mathbf{k}})|\,p^\mu|\phi({\mathbf{p}})\rangle$
of particle 1 and vice versa. 
In more classical terms: the curving of tangential space-time is proportional 
to the distribution of matter, 
if we consider particle 2 as a test-particle and particle 1 as a 
representative of `the rest of the world'. 

This result clearly tells us that there is no uniform curvature, unless
matter is uniformly distributed as in the classical model of de~Sitter 
space-time.

We will now formulate the interaction term $p_\mu c'^\mu$ as an operator in 
Fock space, with $\Phi^\dagger$ and $\Phi$ as emission and absorption 
operators, in the same way as in the case of QED. (A similar procedure is
valid for the other term  $p'_\mu c^\mu$.)
Consider the contribution 
\begin{equation}
\Phi^\dagger ({\mathbf{p}}+{\mathbf{k}})\, p_\mu\, \Phi({\mathbf{p}}) \,\, 
\Phi^\dagger ({\mathbf{p}}'-{\mathbf{k}})\, c'^\mu\, \Phi({\mathbf{p}}').
\label{7-7} 
\end{equation}
By collecting all terms that contribute to given ${\mathbf{p}}$ and 
${\mathbf{k}}$ we obtain
\begin{equation}
\Phi^\dagger ({\mathbf{p}}+{\mathbf{k}})\, p_\mu\, \Phi({\mathbf{p}}) \,\, 
\Gamma^\mu({\mathbf{k}})
\label{7-8} 
\end{equation}
with
\begin{equation}
\Gamma^\mu({\mathbf{k}}) := \int dV(p')\, \Phi^\dagger 
({\mathbf{p}}'-{\mathbf{k}})\, c^\mu\, \Phi({\mathbf{p}}').
\label{7-9} 
\end{equation}
If we replace $\Gamma^\mu({\mathbf{k}})$ by a potential 
$G^\mu({\mathbf{k}})$, in the same way as we have introduced 
$A^\mu({\mathbf{k}})$, we obtain from (\ref{7-8}) - at least formally - 
an interaction term in space-time 
\begin{equation}
\Phi^\dagger (x)\, \partial_\mu\, \Phi(x) \,\times\, G^\mu(x).
\label{7-10} 
\end{equation}
This operator can - in principle - be used as an interaction term within 
a properly defined perturbation algorithm in a 5-dimensional 
pseudo-Euclidean space.

Within our picture of `test-particle' and `massive body' we come to the 
following interpretation of this interaction term. 
$G^\mu(x)$ describes a potential that acts on the test-particle 
in that it adds to its momentum in a way that curves space-time. 
The magnitude of this potential is controlled by the distribution of 
4-momentum of the massive body as a multiplicative factor.
Of course, the test-particle also acts on the massive body by the  
other term $p'_\mu c^\mu$. Therefore, this statement can be generalized 
to: \begin{em}curving of space-time is proportional to the distribution of 
energy-momentum\end{em}.

If there were a well-established theory of quantum gravity, we would now 
proceed and try to bring this interaction term into a form that can be 
compared with the established theory - as we have done in the case of QED.
But at present, we only have at our disposal a theory of gravitation in the
classical domain.
Therefore, we will find out what the structure of this interaction term can 
tell us about the properties of a classical limit to this quantum mechanical 
system. 
This will give us a hint whether or not our model is suited to describe a 
realistic gravitation-like interaction.

\section{Classical correspondence limit}

We already have found that the interaction term causes a curving of 
space-time, and that this curving is not uniform as one might expect if 
one has the cosmological model of `de~Sitter space-time' in mind. 
Instead it is proportional to the distribution of 4-momentum. 
If we start from a flat space-time then the interaction will cause
modifications of the metric tensor. The central question is:
what are the parameters, in the classical limit, that these 
modifications depend on? 

Obviously, the curving does not depend on any internal property of the 
test-particle but is directly linked as an additive term to its momentum, 
thereby revealing an `universal' character of the curving mechanism.
The property of universality gives us a good chance that we can express 
the bending of space-time that our test-particle experiences by a 
modification of the metric tensor that is equally valid for any other 
test-particle. 
This then leads us to the concept of a non-Euclidean space-time with its
curvature determined by the distribution of matter.

Our concept is invariant with respect to the symmetry operations of SO(3,2).
Therefore, it must be possible to find a formulation of a classical limit
that is covariant with respect to the operations of SO(3,2). 
We know that with respect to the neighborhood of any given point these 
operations can be understood as the application of the homogeneous Lorentz 
group (as a subgroup of SO(3,2)) and, in addition, of four operations that 
can be approximated by translations under certain conditions, but, in 
general, include a bending of space-time. 
From this we conclude, firstly, that the relation between curvature and
distribution of matter must be covariant with respect to the homogeneous 
Lorentz group. 
And, secondly, that covariance has to be extended to a non-Euclidean
metric, which means general covariance in the sense of classical general 
relativity.

We can understand the interaction term (\ref{7-10}) as a description of
the variation of the metric tensor, due to the interaction. Then the
form of (\ref{7-10}) tells us that the variation at the point $x$ is 
related to the amount of matter \begin{em}at the same point\end{em} $x$.
This means that the relation between curvature and matter may contain the
metric tensor itself and also its derivatives at the point $x$. 

As long as we restrict ourselves to `small' curvatures, it will be
sufficient to consider only such terms that contain no higher than the 
second derivative of the metric tensor and that are linear in the second 
differential quotient. 
As we know from general relativity \cite{ae}, under these conditions there 
is only one tensor with vanishing divergence that can be built from the 
metric tensor, namely 
\begin{equation}
R_{\mu\nu} - \case{1}{2} g_{\mu\nu} R .  \label{8-1}
\end{equation}
(The definitions of $R_{\mu\nu}$ and $R$ can be found in \cite{ae} or any
testbook on general relativity.)

On the other hand, a covariant description of the distribution of 
4-momentum is uniquely given by the energy-momentum tensor $T_{\mu\nu}$. 
Its divergence must vanish because of energy-momentum conservation.

Then the properties of the interaction term, as discussed above, determine
that the curvature-generating mechanism must be proportional to the
energy-momentum tensor.
This leads us to the well-known field equations of general relativity 
\begin{equation}
R_{\mu\nu} - \case{1}{2} g_{\mu\nu} R = - \kappa T_{\mu\nu}    \label{8-2}
\end{equation}
with an unknown `coupling constant' $\kappa$.

\section{Current-current coupling}

In deriving expression (\ref{2-1}) we had collected all contributions
of magnitude $p^2$ and $p^1$. We had neglected contributions of the 
form $\gamma^\mu \gamma'_\mu$ because they are of magnitude $p^0$ and,
therefore, expected to be very weak. 
Such terms are of the same order of $p$ as $j^{\mu\nu} j_{\mu\nu}$,
which under transformations of the Poincar\'{e} group behaves like a
contribution to a mass term.
This prompts the question whether under these circumstances we can still 
consider the latter as constant. We do not know the answer yet.  

In any case, there may be situations where we are able to observe 
contributions of $\gamma^\mu \gamma'_\mu$.  
So let us take up this term now and consider its contribution 
\begin{equation}
{\bar{b}({\mathbf{p'}}) \, \gamma_\mu \, } {b({\mathbf{p''}})\,\,} 
{\bar{b} ({\mathbf{p'''}}) \, \gamma^\mu} \, b(\mathbf{p''''}) \label{10-1}
\end{equation}
to the corresponding interaction term in Fock space.

Remembering how we have derived QED, we will replace
a pair of emission and absorption operators by operators of a bookkeeping
system or potentials, respectively, in order to linearize the interaction
term. 
Let us rename the operators of the first particle by $e$ and of the second 
by $\mu$ and then replace two operators in (\ref{10-1}) by 
potential operators with the names $\nu_e$ and $\nu_\mu$.
With this we modify (\ref{10-1}) to 
\begin{equation}
{\bar{e}({\mathbf{p'}}) \, \gamma_\mu \, } {\nu_e({\mathbf{p''}})}\,\,\,  
{\bar{\nu_\mu} ({\mathbf{p'''}}) \, \gamma^\mu} \, \mu(\mathbf{p''''}).  
\label{10-2}
\end{equation}
This means, we have split the interaction term into two parts that are
connected by potentials as causal links.
Notice that the analogue to the integral over $p'$ in (\ref{4-5}) is 
now the sum over the four spinor components in (\ref{10-2}).

We can specialize (\ref{10-2}) to a situation where the second part of
the interaction term is evaluated at a large distance from the first part. 
Then, as we have learned from QED, the linking quanta of the potential become
mass-less quanta taking care of the exchange of spin (and momentum) in this 
case. 
For mass-less spinors the four spin components decouple into two-component 
states of left- and right-handed helicity. 
We identify these states with particles of positive energy,
running forward in time, and anti-particles of negative energy, running 
backward in time, respectively.
These can be separated by projection operators (compare (\ref{1-g}) for 
$\gamma_5$ in Weyl representation)
\begin{equation}
\case{1}{2}\,(1-\gamma_5)\,\,\,\,\mbox{and}\,\,\,\,\case{1}{2}\,(1+\gamma_5) 
\end{equation} 
from the four-component states.
The potentials in (\ref{10-2}) are linked, by construction, to particles. 
Therefore, we insert a projection to `particles', which have a left-handed 
helicity, into (\ref{10-2}).
We then obtain a basic building block in the well-known form of weak 
interaction (we have dropped a trivial factor of $1/4$)
\begin{equation}
{\bar{e}({\mathbf{p'}}) \, \gamma_\mu \, } 
(1+\gamma_5)\, {\nu_e({\mathbf{p''}})}\,\,\, 
{\bar{\nu_\mu} ({\mathbf{p'''}}) \, \gamma^\mu}  
(1+\gamma_5)\, {\mu(\mathbf{p''''})}.  
\label{10-3}
\end{equation}
(We have used the commutation relations of the $\gamma$-matrices to bring
the projection operator of the myon-term to the same position as in the
electron-term.) 
This contribution describes the conversion of a myon and an electron-neutrino 
to an electron and a myon-neutrino. 
It implies a `maximum parity violation' of the interaction provided by this 
term.

Let us now replace the electron-neutrino in (\ref{10-3}) by an 
anti-electron-neutrino, which according to our understanding has a 
right-handed helicity and negative energy. 
In the standard treatment of fermion fields, states with negative energy
running backward in time are handled as anti-particles with positive energy
running forward in time. This reinterpretation implies a reflection with
respect to time. 
The operation of time reflection applied to a Dirac spinor is given by
\begin{equation}
T:\,\, x \longrightarrow x' = (-x_0, {\mathbf{x}}), \,\,\,
\psi(x) \longrightarrow \psi'(x') = \pm \gamma_0 \psi(x) 
\end{equation}
(see e.g. \cite{gs}). Since $\gamma_5$ anticommutes with $\gamma_0$, time 
reflection, therefore, means a replacement of right- by left-handed 
helicity. 
So again only left-handed states are found in this contribution if
we relate it to the standard formulation of weak interaction.
This contribution describes the decay of a myon to an electron, an 
anti-electron-neutrino and a myon-neutrino. 

Although we do not know yet what causes the difference between $e$ and $\mu$,
we can identify a difference between $\nu_e$ and $\nu_\mu$ in the following 
respect.
Since $\nu_e$ and $\nu_\mu$ represent far distant $e$ and $\mu$, respectively,
they have \begin{em}different\end{em} properties in so far as $\nu_e$ links 
only to $e$ and $\nu_\mu$ only to $\mu$. 
This reflects the empirical difference of electron- and myon-neutrinos.

We had good reasons to restrict our consideration to massless quanta.
With a neutrino we connect the idea of a \begin{em}free\end{em} particle, 
which means a quantum that is exchanged over a large distance. As such 
it must have a zero mass. 
Also, because of the weakness of the interaction, there is hardly a chance to 
observe a neutrino that is absorbed immediately after its emission.

\section{Compound states}

So far we have studied only processes where two single-particle states
within $\mathcal{H_O}$ combine to form a two-particle state 
and eventually disintegrate again.

If we consider compound states there is no reason to expect that their 
constituent single-particle states are always locatable in tangential 
space-time, if only the resulting compound state can be located within 
$\mathcal{H_O}$. 

How can we describe states that do not belong to $\mathcal{H_O}$
from the view and in the language of an observer in tangential space-time?

Consider the operator $l_{04}$, the SO(3,2) counterpart of the energy or 
time-translation operator $p_0$. 
It generates a rotation in the SO(3,2) Hilbert space $\mathcal{H}$ within
the 0-4-plane. 
A rotation by $\pi/2$ transforms $l_{i4}$ into $l_{i0}$, $i=1,2,3$, but 
leaves $l_{04}$ invariant. 
Therefore, $l_{i0}$ together with $l_{40}$ can generate a kind of shadow 
tangential space-time with Hilbert space $\widetilde{\mathcal{H_O}}$ in the 
same way as $l_{i0}$ generate the normal one.
$l_{ik}$ still generate space-like rotations but the meaning of momentum and 
boost operators are interchanged. 
States of $\widetilde{\mathcal{H_O}}$ can be labeled by eigenvalues of 
operators $q_i$ that are obtained by applying a contraction limit to $l_{i0}$ 
using condition (\ref{1-1}), in a form adapted to $\widetilde{\mathcal{H_O}}$.

The interesting point is that the energy operator of these states is the
same as of states in $\mathcal{H_O}$. Therefore, the shadow states will 
contribute to the total energy and should thereby add to gravitation. 
On the other hand, they cannot participate in the electromagnetic interaction 
with $\mathcal{H_O}$, because their operators $q_i$ cannot 
exchange `momentum' with the operators $p_i$ of $\mathcal{H_O}$ - at least 
not by the mechanism that we have identified as electromagnetic interaction.  
It is tempting to regard these states as a kind of `dark matter'. 

Consider now special types of states that belong neither to $\mathcal{H_O}$ 
nor to $\widetilde{\mathcal{H_O}}$ nor to any linear combination of regular 
and shadow states. 
We define such states by exchanging one or two momentum components $p_i$ by 
the corresponding $q_i$.

This leads to the following configurations:
$(p_1,p_2,q_3)$, $(p_1,q_2,p_3)$, $(q_1,p_2,p_3)$ and
$(p_1,q_2,q_3)$, $(q_1,q_2,p_3)$, $(q_1,p_2,q_3)$.
We can easily convince ourselves that $p_i$ commute with $q_k$ if $i\not= k$,
by using the commutation relations of SO(3,2) together with (a generalized 
form of) condition (\ref{1-1}). 
This means that the operator triplets deliver as good quantum numbers as the 
$p_k$ of regular states, despite the somewhat coarse way of their 
construction. 
It would be hard to understand if such states were not to be occupied in the 
same way as states of $\mathcal{H_O}$ or $\widetilde{\mathcal{H_O}}$.
Therefore, we have to conclude that such states are an inevitable and 
integral part of the SO(3,2) model.

These states cannot be localized in tangential space-time because one or two of 
the momentum components are missing. 
Therefore, these states do not belong to any irreducible representation of the 
Poincar\'{e} group, which means that they cannot appear as free particles.
But since these states have one or two regular momentum components they may
be able to exchange momentum with regular particles by electromagnetic 
interaction. 
Remembering the factors that contribute to the estimate of the 
electromagnetic coupling constant we can, at least formally, ascribe charges 
of $1/3$ and $2/3$ of $e$ to these states.
 
If we intend to construct compound states that shall appear as particles in 
the tangential space-time, we have to combine the individual states in such a 
way that either the $q_i$ compensate or add up to a symmetric configuration, 
so that the compound state corresponds finally to a representation of the 
Poincar\'{e} group. 

Similar properties has been found empirically within the quark model and 
have triggered the formulation of quantum chromodynamics.
We, therefore, tend to label the three modifications within each of the two 
groups by `color' quantum numbers, and identify the first group with 
`up quarks' and the second with `down quarks'.

Let $a^\dagger_r, a_r, a^\dagger_b, a_b, a^\dagger_g, a_g$ be creation and
annihilation operators of quarks with colors $r, b, g$. Then
the operators 
\begin{eqnarray}
T_+ &=& a^\dagger_r a_b, \,\,\, T_- = a^\dagger_b a_r \nonumber \\
B_+ &=& a^\dagger_r a_g, \,\,\, B_- = a^\dagger_b a_g  \\
C_+ &=& a^\dagger_g a_b, \,\,\, C_- = a^\dagger_g a_r \nonumber 
\end{eqnarray}
exchange the members of a triplet.
It can be shown that together with the counting operators 
\begin{eqnarray}
B &=& a^\dagger_r a_r + a^\dagger_b a_b + a^\dagger_g a_g \nonumber \\
T &=& \case{1}{2}(a^\dagger_r a_r - a^\dagger_b a_b) \\
N &=& \case{1}{3}(a^\dagger_r a_r + a^\dagger_b a_b - 2a^\dagger_g a_g) 
\nonumber  
\end{eqnarray}
they form the Lie algebra of SU(3) \cite{hjl}.

Since none of the colors are preferred to the others, any interaction term 
that involves these states must be invariant with respect to the exchange
operations of SU(3). This means that a generalization of the interaction 
term of QED to compound states built from `quarks', will lead to an 
interaction term of a similar structure as that of quantum chromodynamics. 

As mentioned above, a rotation generated by $l_{04}$ transforms $l_{i4}$ 
into $l_{i0}$, $i=1,2,3$, and vice versa. 
$l_{04}$ is the \begin{em}exact\end{em} SO(3,2) counterpart of 
time-translation. 
Therefore, compound states that are built up from contributions of $p_i$ and 
$q_i$, in a way that is symmetric with respect to operations of $l_{04}$, 
can be stable over an extremely long period of time.
This supports the suspicion that `quark' states play an essential role in the
formation of compound states.

We have sketched here a possible correspondence to QCD in order to illustrate 
that very promising structures can be found within compound states. 
These deserve a deeper evaluation, but such an evaluation is beyond the scope 
of the present article.

\section{Spin foams revised}

Our perturbative approach to the SO(3,2) spin foam model has led us to
familiar structures of perturbative quantum field theory. 
In reverse, we should feel entitled to regard the corresponding sums over 
Feynman graphs as a perturbative description of our spin foam model. 
This gives the `vertices' of the spin foam model a precise meaning, which
allows accurate numerical calculations.
The `edges' of spin foams then have to be identified as incoming and outgoing 
lines of Feynman graphs and are, therefore, labeled by spin-1/2 
representations of the Poincar\'{e} group instead of SO(3,2). 
Internal lines of Feynman graphs corresponding to contraction functions 
have to be considered as part of the perturbation algorithm and should not 
be related to edges of spin foams.

A similar description of spin foams by Feynman graphs has been proposed 
by M. Reisenberger and C. Rovelli \cite{rr}.

\section{Quantum cosmology}

We have found strong evidence for a fundamental symmetry with the 
structure of the group SO(3,2), with the homogeneous Lorentz group as a
subgroup and the full Poincar\'{e} group serving as an approximate symmetry. 
The QED part of our model verifies the SO(3,2) symmetry from subatomic to 
macroscopic scales, whereas the gravitational part has the potential to 
verify the model at least up to the borders of our solar system, and
possibly up to cosmologic scales.

The model consistently forms the structure of space-time by the gravitational 
part and delivers a means to probe space-time by long range photons. 
This should make it an useful quantum mechanically based cosmologic model 
reaching from subatomic to cosmologic scales. 

In studying compound systems we have noticed that there is another tangential
space orthogonal to the first. Both spaces are connected by the energy 
operator, and time \begin{em}rotates\end{em} rather than 
\begin{em}shifts\end{em} one space into the other.
This brings up an interesting point: By randomly selecting a tangential 
space-time the particle world is divided, in principle, into three kinds of 
matter: \\
a) states that are located in the tangential space-time (matter described by
presentations of the Poincar\'{e} group acting in $\mathcal{H_O}$), \\
b) states that are located in the orthogonal shadow space (dark matter), \\
c) states that are defined by mixed quantum numbers of both spaces (quarks).\\
This defines a mechanism of symmetry breaking caused by using tangential 
space-time as the basis of the theory.

Of course, there are states that belong to tangential spaces between those of 
case a and b. 
Concerning these states, the electromagnetic part probably has to be 
generalized for the exchange of photons over very long distances back in time. 
There the rotation from $\mathcal{H_O}$ towards $\widetilde{\mathcal{H_O}}$ 
can become noticeable, and the model will clearly guide us how to accomplish 
this generalization. 

There has been some concern that the SO(2) subgroup of rotations in the 
0-4-plane could lead to unphysical time loops (Grandfather paradox).
In fact, the operator $l_{04}$ corresponding to time translations generates 
rotations of the Hilbert space $\mathcal{H}$. However, no experiment
would be able to observe a rotation of the \begin{em}entire\end{em} 
$\mathcal{H}$. 
Therefore, the $l_{04}$ formally generates a kind of periodic background time,
but to an observer it appears like a translation with an infinite range. 
An appropriate picture of the structure of this background time is that of a 
helix rather than a circle.

Nevertheless, there are means to observe effects of a rotation if we probe
space-time by observing photons from events at very large distances back in 
the past, as mentioned before. 
This may show the observed red shift of distant galaxies in a new light.
 
Background time is useful for embedding events that are defined as transitions 
between quantum states. 
But it is the sequence of events that forms our imagination of time and that 
finally defines the physical properties of time. 
There is definitely no periodicity connected with a long sequence of 
statistical events. Especially, within our model the grandchild will never 
have a chance to meet his grandfather in a time loop.

\section{Conclusion }

We have established a model of space-time and of particle physics therein 
that in some sense is parallel to the standard model in that it describes 
interactions from a different point of view.
However, this model has proven much more stringent in that it         \\
- delivers models of all four known types of interactions (and only these) 
by the same symmetry principle,                                       \\
- allows to determine coupling constants,                             \\
- delivers a particle spectrum consisting of massive leptons (so far
we do not know what makes up the difference between electron, myon and
tau), 
photons, neutrinos, quarks and compound states made up by quarks,     \\
- explains gauge invariance (in case of QED) and SU(3) invariance (in case of 
QCD) and violation of parity (in case of weak interaction),           \\
- describes the formation of space-time as a combined action of electromagnetic
and gravitational interaction.

Space-time has formally been obtained as tangential space-time by contracting 
the SO(3,2) symmetry group to the Poincar\'{e} group. 
This does not come as a surprise since this is a consequence of the 
representations that we have chosen.
However, in connection with the properties of the electromagnetic interaction 
term, which is `local' in space \begin{em}and time\end{em}, it adopts 
qualities of an observable physical space-time continuum: the interaction term 
allows - in principle - to perform measurements of particles at `points' 
in space and time. 
This realized the notion of \begin{em}events in space and time\end{em}.

Also the auxiliary `photon' field obtains `real' physical properties identical
to the empirical electromagnetic field in the sense that the interaction term 
allows us to probe its action on `charged particles' at different points of 
space-time.
In addition, photons represent \begin{em}causal links\end{em} connecting
events that are caused by the interaction term.

All results described in this article have been obtained by replacing the 
Poincar\'{e} symmetry of the standard model by a de~Sitter symmetry. 
Except for the presumption of large quantum numbers no additional assumptions 
like gauge invariance or `higher' symmetries have been made. 

The implications of this replacement for particle physics makes it very likely
that the de~Sitter group SO(3,2) defines a fundamental symmetry of nature, 
well-hidden by innocent-looking interaction terms. Only to the extend that 
interactions can be neglected it can be approximated by the Poincar\'{e} 
group.

\renewcommand{\baselinestretch}{1.1}


\begin{thebibliography}{99}

\bibitem{rp} R.~Penrose, ``Angular momentum: an approach to 
combinatorial space-time", in: \begin{em}Quantum Theory and Beyond\end{em},  
ed. Ted Bastin (Cambridge University Press, Cambridge, 1971).

\bibitem{jcb} J. C. Baez, ``An introduction to spin foam models of
quantum gravity and BF theory", in: \begin{em}Geometry and Quantum 
Physics\end{em}, eds. H. Gausterer and H. Grosse, (Springer, Berlin, 2000).
Available as gr-qc/9905087.

\bibitem{rh1} R. Haag, 
Nucl. Phys. B (Proc. Suppl.), \bfseries 18B\mdseries, 135 (1990).

\bibitem{rh2} R. Haag, \begin{em}Local Quantum Physics\end{em}, 
(Springer, Berlin Heidelberg New York, 1992 and 1996).

\bibitem{fmls} F. Markopoulou and L. Smolin, 
Nucl. Phys. \bfseries B508\mdseries, 409, (1997).

\bibitem{fm} F. Markopoulou, ``Dual formulation of spin network evolution", 
gr-qc/9704013.

\bibitem{aw} A. Wyler, 
C. R. Acad. Sc. Paris \bfseries269\mdseries, 743 (1969).

\bibitem{iw} E. In\"on\"u and E. P. Wigner, 
Proc. Nat. Acad. Sci. USA \bfseries 39\mdseries, 510 (1953).

\bibitem{fg} F. G\"ursey in: \begin{em}Group Theoretical Concepts and
Methods in Elementary Particle Physics\end{em}, ed by G\"ursey  
(Gordon and Breach, 1964). 

\bibitem{ff} E. Angelopoulos, M. Flato, C. Fr{\o}nsdal, and D. Sternheimer, 
Phys. Rev. D \bfseries 23\mdseries, 1278 (1981). 

\bibitem{gs} G. Scharf, \begin{em}Finite Quantum Electrodynamics\end{em}, 
(Springer, Berlin Heidelberg New York, 1989 and 1995).

\bibitem{rf} R. Feynman, Phys. Rev. \bfseries 76\mdseries, 749 (1949). 

\bibitem{pam} P.A.M. Dirac, \begin{em}The Principles of Quantum 
Mechanics\end{em}, (Oxford University Press, Oxford, 1958).

\bibitem{hl} L. K. Hua and K. H. Look (= Lu, Qi-keng),
Scientia Sinica \bfseries8\mdseries, 1031 (1959).

\bibitem{fkklr} J. Faraut, S. Kaneyuki, A. Kor\'{a}nyi, Qi-keng Lu, G. Roos, 
\begin{em}Analysis and Geometry on Complex Homogeneous Domains\end{em},
(Birkh\"auser, Boston Basel Berlin, 2000).

\bibitem{lkh} L. K. Hua, \begin{em}Harmonic Analysis of Functions of
Several Complex Variables in the Classical Domains\end{em},
(American Mathematical Society, Providence, 1963).

\bibitem{br} B. Robertson, 
Phys. Rev. Lett. \bfseries27\mdseries, 1545 (1971).

\bibitem{fds} F. D. Smith, Jr., 
Int. J. Theor. Phys. \bfseries24\mdseries, 155 (1985);
\bfseries25\mdseries, 355 (1986). 

\bibitem{kino} T. Kinoshita, preprint, Conference on Precision Electromagnetic
Measurements, CLNS96/1418 (1996).

\bibitem{ae} A. Einstein, \begin{em}Grundz\"uge der Relativit\"atstheorie
\end{em}, p.54, (Friedr. Vieweg und Sohn, Braunschweig, 1960).

\bibitem{hjl} H. J. Lipkin, \begin{em}Lie Groups for Pedestrians\end{em}, 
p.33, (North-Holland Publishing Company, Amsterdam, 1965).

\bibitem{rr} M. Reisenberger, C. Rovelli, 
``Spin foams as Feynman diagrams", gr-qc/0002083.

\end{thebibliography}
\end{document}